\documentclass[ reprint,  amsmath,amssymb,  aps, ]{revtex4-1}
\usepackage[dvipdfmx]{graphicx}
\usepackage{dcolumn}
\usepackage{bm}
\usepackage{amsmath}
\usepackage{amsfonts}
\usepackage{amssymb}
\usepackage{color}
\usepackage{comment}
\providecommand{\U}[1]{\protect\rule{.1in}{.1in}}
\begin{document}

\preprint{APS/123-QED}

\title{Grand Projection State: A Single Microscopic State to Determine Free Energy}
\author{Tetsuya Taikei}
\author{Tetsuya Kishimoto}
\author{Kazuhito Takeuchi}
\author{Koretaka Yuge}
\affiliation{Department of Materials Science and engineering, Kyoto University, Kyoto 606-8501, Japan}

\begin{abstract} 
Recently, we clarify connection of spatial constraint and equilibrium macroscopic properties in disordered states of classical system under the fixed composition; namely few special microscopic states, independent of constituent elements, can describe macroscopic properties.
In this study, we extend our developed approach to composition-unfixed system. Through this extension in binary system, we discover a single special microscopic state to determine not only composition but also Helmholtz free energy measured from unary system, which has not been described by a single state.
\end{abstract}
\maketitle

\section{\label{sec:level1}Introduction}
It has been more than 130 years since Hermann von Helmholtz defined free energy $F$. $F$ is very useful concept when we can consider the system to be under constant volume, like crystalline solid and liquid. $F$ has the member of entropy, which is a measure of the number of possible microscopic states. Therefore, $F$ has not been described by a single state in classical system.
\\
\ In statistical mechanics, free energy and macroscopic physical properties ($\overline{C}$) can be given by
\begin{eqnarray}
F&=&-k_{\rm{B}}T\ln{Z},\label{eq:ceproper1} \\
 \overline{C}&=&Z^{-1}\sum_{d} C^{(d)}\exp\left(-\frac{E^{(d)}}{k_{\rm{B}}T}\right).
\end{eqnarray}
Here,  $d$ is a microscopic state on phase space,
 $k_{\rm{B}}$ is Boltzmann constant, $T$ is temperature,
and $Z$ is partition function.
$Z$ takes sum over possible microscopic state, thus $F$ cannot be described by a single state from Eq.~\ref{eq:ceproper1}. It is obvious that with increase of system, the number of possible states increases exponentially. This fact makes direct estimation of $F$ practically impossible.
 Therefore, several calculation techniques have been developed and widely utilized, such as entropic sampling {\cite{EntropicSampling} with Monte Carlo  (MC) simulation {\cite{MonteCarlo1}\cite{MonteCarlo2}},
cluster variation method\cite{CVM1}\cite{CVM2}, and Frankel method\cite{Frankel}.
\\
\ Recently, we clarify relationship between spatial constraint (viz. lattice for crystalline solids, volume and density for liquid) and equilibrium macroscopic properties in disordered states of classical system under the fixed composition\cite{EMRS1}\cite{EMRS2}\cite{EMRS3}.
Here, we do not need temperature, elements constituent, or interactions.
This approach relies on that density of state for $C$ on configuration space can be well characterized by multidimensional Gaussian distribution.
This important characteristic of distribution of $C$  is proven by combination of a theory and an approximation.
\\
\ The theory is that microscopic states on configuration space can be described by complete basis functions \{${q_1,q_2,....q_s}$\}. 
By using basis functions, $C$ for any microscopic state $d$ can be expressed as
\begin{equation}
	C^{(d)}=\sum_{r=1}^{s} {\langle}C{\mid}q_r\rangle q_r^{(d)},
\label{eq:CEM}
\end{equation}
where ${\langle}C{\mid}q_r\rangle$ means inner product. Thus, when $C$ is energy and $q_r$ is cluster function\cite{CEM1}\cite{CEM2}\cite{CEM3},
${\langle}C{\mid}q_r\rangle$ denotes effective cluster interaction, called ECI.
The approximation is that
the density of states for $q_r$ on non-interacting system is expressed by Gaussian distribution.\cite{EMRS3} The validity of this approximation is confirmed by random matrix research; this means vanishment of statistical interdependence of $q_r$\cite{EMRS4}. These two points shows that density of state for $C$, linear combination of $q_r$, is well characterized by multidimensional Gaussian distribution.  
\\
\ This developed approach enable us to estimate free energy directly, and to get few special microscopic states, which determine $\overline{q_r}$ and $\overline{C}$. However, we have not been successful in considering composition and describing free energy by one state even with our approach.
\\
\ In the present study, we extend our developed approach under fixed composition (canonical approach) to composition-unfixed system (semi-grand canonical approach). In semi-grand canonical approach, we consider the system with the framework of semi-grand  canonical ensemble, thus chemical potential is needed.
Unlike grand canonical ensemble, semi-grand one needs number of all atoms.
This extension is suggested by additional research with random matrix.
Through this study, we get a single special microscopic state in binary system of disordered states under classical system, which is derived only from spatial constraint. We call this state \\``{Grand Projection State}'' (GPS).
%
%
GPS give us not only composition in the system, but also Helmholtz free energy $F$ measured from unary system, which has not been represented by a single state in classical system. We explain additional random matrix research, and introduce present approach to get composition and free energy with GPS.
 \section{\label{sec:level2}Derivation and concept}
 \subsection{\label{sec:level1} Random Matrix}
 Let us first explain our random matrix (RM) approach. This approach is introduced for confirming the characteristic of  density of microscopic states, and we show that statistical interdependence of $q_r$ becomes disappeared as the scale of spatial constraint becomes large\cite{EMRS4}.
 \\
 \ Here, we take $m \times n$ matrix $K$, where $m$ is the number of sampling point on configuration space, and $n$ is the number of basis functions considered. With this definition, the normalized covariance matrix $J$ for $q_r$can be calculated as
 \begin{equation}
 J=\frac{1}{m}K^{\rm{T}}K,
 \label{eq:matrix}
\end{equation}
where  $K^{\rm{T}}$ means transposed matrix of $K$. 
When we consider an ideal system where statistical interdependence is disappeared (viz. $K$ is RM, $K_{\rm{RM}}$), all elements  is independently consisted of normal random numbers, with its average and variance respectively taking 0 and 1, leading to all diagonal elements for RM, $J_{\rm{RM}}$, to be 1. The validity of constructed RM in this research with finite size is guaranteed, comparing with Marchenko-Pastur distribution\cite{Marchenko}.
Meanwhile, when we consider practical system, we first uniformly sample $m$ microscopic states using MC simulation. Then, we calculate and normalize the value of $n$ basis functions in each microscopic state so that average and variance of each column of $J$ respectively should be 0, and $1/m$ for comparison.  When the statistical interdependence of $q_r$ is disappeared, the elements of matrix $J$ is regarded to be consisted independently, which is same as RM. Therefore, we check the difference of matrices using their eigenvalues, as density of eigenvalues (DOE) of covariance matrix $J$ .
%
For setting practical system, we have two choices; composition-fixed method and composition-unfixed method (see appendix A).
\\
\ In this study, we consider an example for equiatomic A-B binary system on FCC lattice. In order to get basis function, we employ generalized Ising-like spin of $\sigma={\pm}1$, and $q_r$ can be defined by $q_r=\langle\prod_{i{\in}k} \sigma_i\rangle_{\rm{lattice}}$, where $\sigma_i$ is spin at site $i$, $\langle\cdot\cdot\rangle_{\rm{lattice}}$ is average over all sites on the lattice, and $k$ is the index indicating cluster type, such as empty, point, 1st nearest neighbor (1NN) pair. Advantage of applying Ising-model is that the interdependent of $q_r$ has already been none-zero with taking limit of the number of atoms in the system without a change of basis\cite{EMRS3}.
Here, clusters considered are pair clusters up to 6NN, triplet clusters consisting of up to 6NN pairs, resulting in 29 basis functions(i.e. $n$=29). We sample 500,000 microscopic states (i.e. $m=500,000$), and perform 500,000 MC steps for 1152-atom MC-cell for getting one microscopic state. We can see that statistical interdependence is more vanished in composition-unfixed practical system (CUFS) than in composition-fixed practical system (CFS).
\begin{figure}[t]
\centering
\includegraphics[width=8cm]{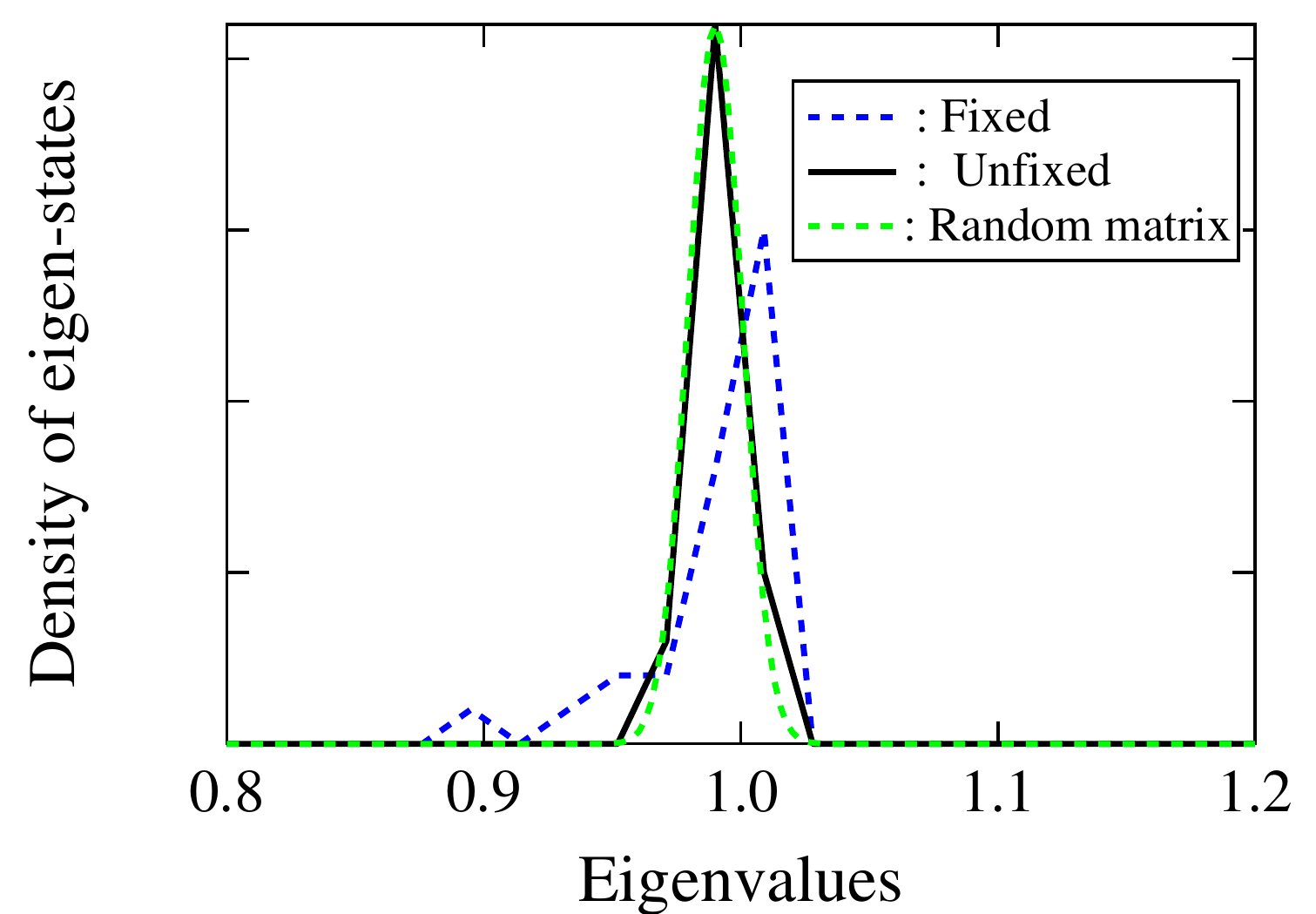}
\caption{Density of states along eigenvalues of covariance matrix (DOE), constructed from  CFS, CUFS, and RM. This landscape shows CUFS is more similar to RM than CFS.}
\label{fig.dos_unfixed}
\end{figure}
\begin{figure}[b]
  \begin{center}
    \begin{tabular}{c}
%
      \begin{minipage}{0.5\hsize}
        \begin{center}
          \includegraphics[width=4.5cm]{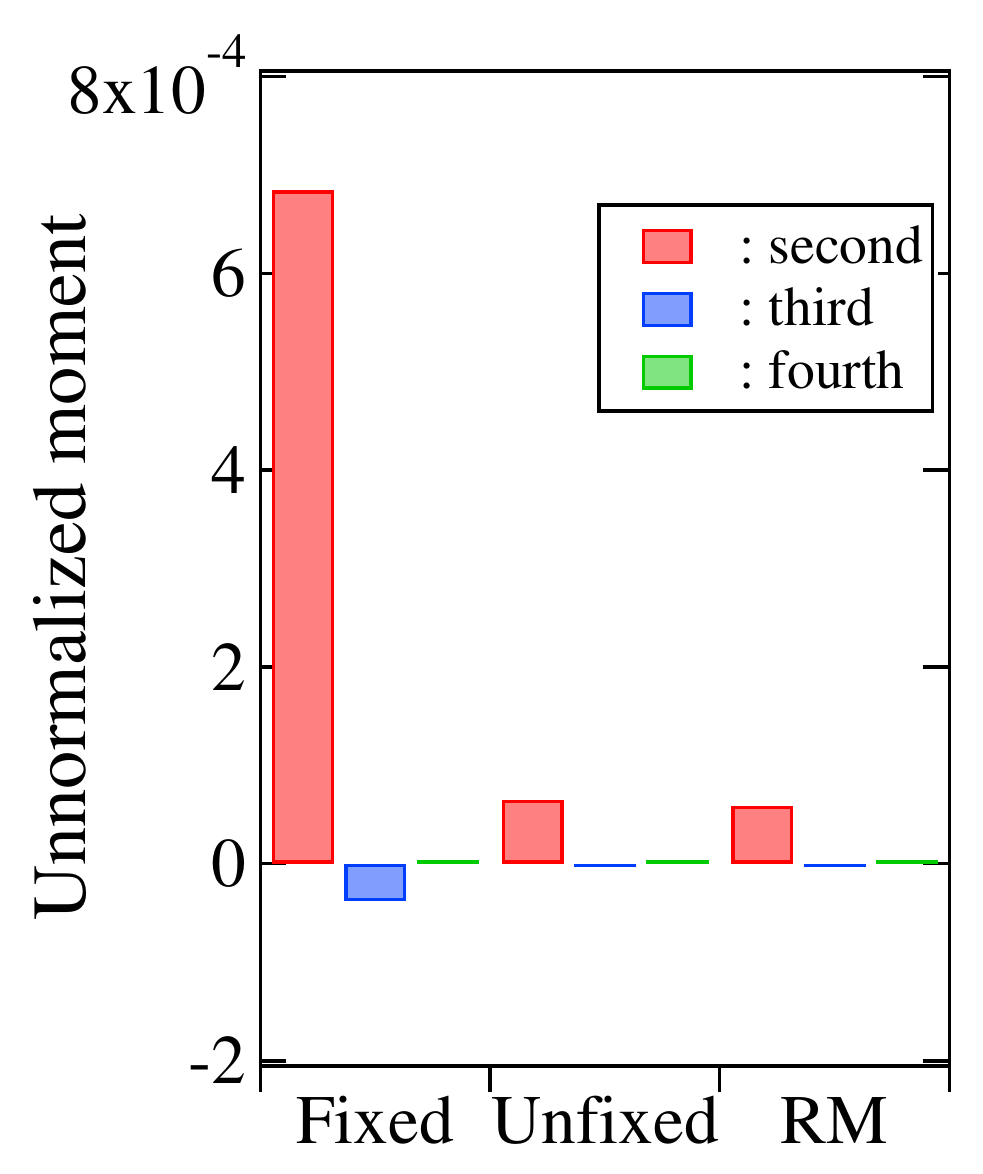}
        \end{center}
      \end{minipage}
      \begin{minipage}{0.5\hsize}
        \begin{center}
          \includegraphics[width=4.5cm]{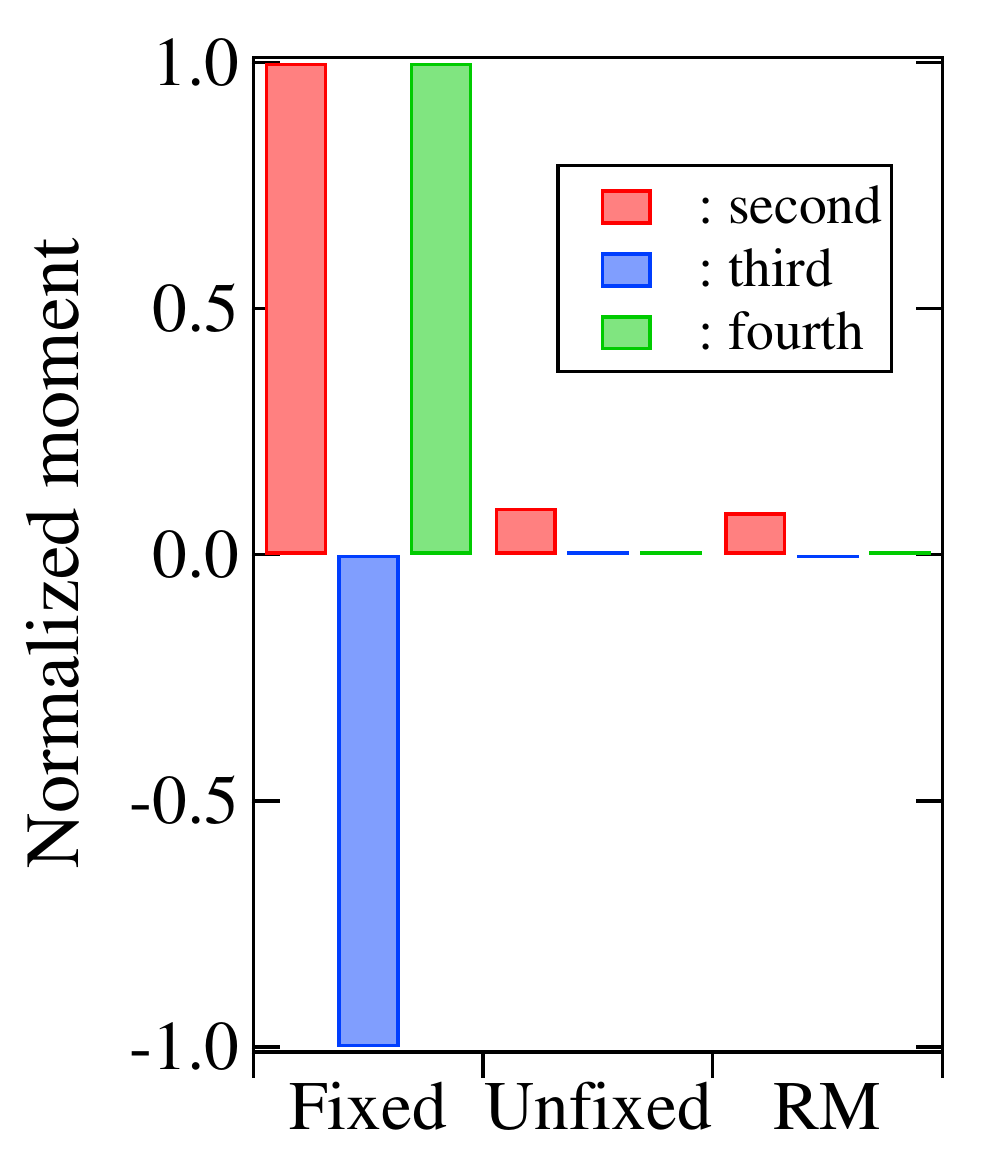}
        \end{center}
      \end{minipage}
    \end{tabular}
    \caption{2nd, 3rd, and 4th-order moment of DOEs for CFS, and CUFS, and RM.
        These shows similarity of RM and CUFS quantitatively.  \\
    Left: The value is unnormalized.\\
    Right: The value is normalized for comparison so that the magnitude of each order moment for CFS is 1.}
    \label{fig.moment_unfixed}
  \end{center}
\end{figure}
\\
\ Figure~\ref{fig.dos_unfixed} shows DOEs for CFS, CUFS, and RM. Although difference between CFS and RM is gradually disappeared when number of atoms increases \cite{EMRS4}, they still have quantitative differences: DOE for RM has single sharp peak, while CFS does not have. Meanwhile for CUFS, under the same number of atoms, it is easily realized that the landscape of DOE is much more similar to RM than CFS in terms of shape and location of the peak. To make further quantitative comparison, we estimate moments (from 2nd to 4th) of DOEs, defined by
\begin{eqnarray}
M_1&=&\frac{\Sigma_{i=1}^{N_d} x_i}{N_d},\\
M_L&=&\frac{\Sigma_{i=1}^{N_d} (x_i-M_1)^L}{N_d},
\end{eqnarray}
where $L$ is an order of the moment, $N_d$ is number of data and $x_i$ is each data of index $i$. 
 Figure~\ref{fig.moment_unfixed} shows moments in  CFS, CUFS, RM. 2nd-order moment of CUFS successfully agree with that of RM.  In addition to 2nd-order, compared to CFS, 3rd and 4th-order is vanished in CUFS, which shows excellent agreement with RM. 
\\
\ These fact shows that CUFS is much more similar to RM, namely statistical interdependence is more vanished in CUFS than in CFS. Therefore, composition should be considered in setting practical system, and this suggests that our established approach in disordered states should be extended to composition-unfixed system; semi-grand canonical ensemble.
We show concept and derivation of the extension in binary system.
 \subsection{\label{sec:level2} Grand Projection State (GPS)}
In canonical ensemble, we have shown that canonical average of basis function $q_r$, $Q_r$, can be expressed by
 \begin{equation}
Q_r(T)\simeq{\langle}q_r\rangle_1\mp\sqrt{\frac{\pi}{2}}{\langle}q_r\rangle_2\cdot\frac{E_{r\pm}^{\rm{proj}}-{\langle}E\rangle_1}{k_{\rm{B}}T},
\label{eq:Projection_CE}
\end{equation}
 \begin{equation}
E_{r\pm}^{\rm{proj}}=\sum_{t=1}^{s} {\langle}E{\mid}q_t\rangle{\langle}q_t\rangle_r^{(\pm)},
\label{eq:Projectionenergy_CE}
\end{equation}
where, $\langle\cdot\cdot\cdot\rangle_1$ and $\langle\cdot\cdot\cdot\rangle_2$ respectively denotes average and standard deviation over all microscopic state  \cite{EMRS2}(and see Appendix B). $\langle\cdot\cdot\cdot\rangle_r^{(+)}$ and $\langle\cdot\cdot\cdot\rangle_r^{(-)}$ is a partial average over all microscopic states respectively under $q_r\ge{\langle}q_r{\rangle}_1$ and $q_r\le{\langle}q_r{\rangle}_1$. Then, $E_{r\pm}^{\rm{proj}}$ respectively can be described by a single microscopic state, \{${\langle}q_1\rangle_r^{(\pm)},{\langle}q_2\rangle_r^{(\pm)},...{\langle}q_s\rangle_r^{(\pm)}$\}, which is derived only from the information of spatial constraint (e.g., lattice). We call this state projection state, and $E_r^{\rm{proj}}$ projection energy along coordination $r$. It is notable that Eq.~(\ref{eq:Projection_CE}) is derived from the fact that density of states on configuration space for non-interacting system can be well characterized by multidimensional Gaussian distribution. Two representation of $Q_r(T)$ comes from this approximation, and when density of states completely matches Gaussian distribution, two values are same.
\\
\ In semi-grand canonical ensemble, instead of $E$, we consider $\Delta{\mu}$ and $I$ defined as
  \begin{equation}
\Delta{\mu}= \mu_{\rm{A}}-\mu_{\rm{B}},
\label{eq:Definition_delmu}
\end{equation}
\begin{equation}
I=E-{\Delta}{\mu}Nx_{\rm{A}},
\label{eq:Definition_I}
\end{equation}
 with Legendre transformation; $E$ to $I$. Here, $\mu_{\rm{A}}$ and $x_{\rm{A}}$ respectively means chemical potential energy and  composition of A atoms, and $N$ means number of all atoms (constant in semi-grand canonical). Then in semi-grand canonical ensemble, we can write the ensemble average like Eq.~(\ref{eq:ceproper1}) as
  \begin{eqnarray}
\overline{C_G} 
&=&AZ_G^{-1}\sum_{d}C^{(d)}\exp(-\frac{I^{(d)}}{k_BT}),
\label{eq:Definition_delmu}
\end{eqnarray}
where $A=\exp({\frac{\mu_BN}{k_BT}})$, and $Z_G$ is sem-grand partition function. We take care about a possible microscopic state $d$ including composition, and we respectively define $\langle\cdot\cdot\cdot\rangle_{1G}$ and $\langle\cdot\cdot\cdot\rangle_{2G}$ as average and standard derivation including composition. Now in binary system, we define coordination for composition, $q_{\rm{comp}}=x_A$. With this definition ${\Delta}{\mu}Nx_A$ can be described by coordination for composition; therefore same as $E$, we can express $I$ with inner product like Eq.~(\ref{eq:CEM}):
\begin{eqnarray}
I^{(d)}&=& \sum_{t=1}^{s} {\langle}E{\mid}q_t{\rangle}q_t^{(d)}-{\Delta}{\mu}N{\cdot}q_{\rm{comp}} \label{eq:product1} \\
 &=& \sum_{t=1}^{s} {\langle}I{\mid}q_t\rangle q_t^{(d)}.
\label{eq:product2}
\end{eqnarray}

    In addition to this, density of states along $q_{\rm{comp}}$ shows Gaussian distribution, because it is from binomial distribution.  Therefore, we can write $I$ as multidimensional Gaussian distribution; thus, we can apply our canonical approach to semi-grand canonical one with minor change. Just we should do is to replace $E$ by $I$; Equations~(\ref{eq:Projection_CE}) and (\ref{eq:Projectionenergy_CE}) can be extended:
 
  \begin{equation}
\grave{Q_r}(T)\simeq{\langle}q_r\rangle_{1G}\mp\sqrt{\frac{\pi}{2}}{\langle}q_r\rangle_{2G}\cdot\frac{I_{r\pm}^{\rm{proj}}-{\langle}I{\rangle}_1}{k_{\rm{B}}T}.
\label{eq:Projection_GCE}
\end{equation}
 \begin{eqnarray}
I_{r\pm}^{\rm{proj}}&=&\sum_{t=1}^{s} {\langle}I{\mid}q_t\rangle{\langle}q_t\rangle_{r}^{({\pm}G)} \nonumber \\
&=&\sum_{t=1}^{s} {\langle}E{\mid}q_t\rangle{\langle}q_t\rangle_r^{({\pm}G)}-{\Delta}{\mu}N{\cdot}{\langle}q_{\rm{comp}}\rangle_r^{({\pm}G)}\nonumber\\
&=&\grave{E}_{r\pm}^{\rm{Proj}}-{\Delta}{\mu}N{\cdot}{\langle}q_{\rm{comp}}\rangle_r^{({\pm}G)}.
\label{eq:Projectionenergy_GCE}
\end{eqnarray}
\begin{equation}
\grave{E}_{r\pm}^{\rm{Proj}}=\sum_{t=1}^{s} {\langle}E{\mid}q_t\rangle{\langle}q_t\rangle_{r}^{({\pm}G)}.
\label{eq:E_semi_grand}
\end{equation}
Here, $\langle\cdot\cdot\cdot\rangle_r^{({\pm}G)}$ denotes partial average including composition,  $\grave{E}$ denotes projection energy in semi-grand canonical ensemble, and $\grave{Q_r}$ denotes semi-grand canonical average of basis function $q_r$. When we think coordination $r$ to be composition, we can get special information and advantage. We call this special projection state for $q_{comp}$, Grand Projection State (GPS). 
First, GPS determine composition in the system. From symmetry of binomial distribution, which leads density of states along $q_{\rm{comp}}$, we can easily introduce ${\langle}q_{\rm{comp}}\rangle_{1G}=1/2$. In addition, $I_{\rm{comp+}}^{\rm{proj}}$ and $I_{\rm{comp-}}^{\rm{proj}}$ is calculated respectively based on A-rich, B-rich composition. 
 Therefore, from Eq.~(\ref{eq:Projection_GCE}), we get two representation of ensemble average of composition (for the sake of simplicity, hereinafter ${\langle}I\rangle_1 = 0$);
 \begin{eqnarray}
 x_{\rm{A}}(T)&\simeq&\frac{1}{2}-\sqrt{\frac{\pi}{2}}{\langle}q_{\rm{comp}}\rangle_{2G} \frac{I_{\rm{comp+}}^{\rm{Proj}}}{k_{\rm{B}}T}.\label{eq:composition}
 \\
  x_{\rm{B}}(T)&\simeq&\frac{1}{2}-\sqrt{\frac{\pi}{2}}{\langle}q_{\rm{comp}}\rangle_{2G} \frac{I_{\rm{comp-}}^{\rm{Proj}}}{k_{\rm{B}}T}.
\end{eqnarray}
Here, we get composition in the system from a single state, GPS, \{${\langle}q_1\rangle_{\rm{comp}}^{(+G)},{\langle}q_2\rangle_{\rm{comp}}^{(+G)},...{\langle}q_s\rangle_{\rm{comp}}^{(+G)}$\}, or \{${\langle}q_1\rangle_{\rm{comp}}^{(-G)},{\langle}q_2\rangle_{\rm{comp}}^{(-G)},...{\langle}q_s\rangle_{\rm{comp}}^{(-G)}$\}, which depends only on spatial constraint. Moreover, GPS can be given analytically unlike the other projection states (see appendix A). \\
\ We consider only phase space above, then we consider momentum space. Our approach relies on Gaussian distribution. When the lifetime of a particular configuration is typically long enough to achieve vibrational equilibrium state, we can represent vibrational free energy \cite{F_vib} by microscopic state on configuration space; namely momentum space can be characterized by Gaussian distribution. Consequently, we can apply our approach not only to crystalline solids, also to liquid and amorphous material. 
\\
Now, free energy $F$ is 
 \begin{eqnarray}
F&=&E-TS+{\mu}N \nonumber \\
&=&E-TS+\mu_{\rm{B}}N+\Delta{\mu}Nx_{\rm{A}}.
\label{eq:Projectionenergy_GCE2}
\end{eqnarray}
Thereby, 
\begin{equation}
\frac{\partial F}{\partial x_{\rm{A}}}=\Delta{\mu}N,\ \ \frac{\partial F}{\partial x_{\rm{B}}}=-\Delta{\mu}N.
\end{equation}
Therefore, we get $F$ measured from unary system;
\begin{eqnarray}
F&=&F_{\rm{A}}+\int^{x_{\rm{A}}(T)}_1\Delta{\mu}N dx_{\rm{A}}.
\label{eq:free_A}\\
F&=&F_{\rm{B}}+\int^{x_{\rm{B}}(T)}_1(-\Delta{\mu}N) dx_{\rm{B}}.
\label{eq:free_B}
\end{eqnarray}
With Eqs.~(\ref{eq:Projectionenergy_GCE})-(\ref{eq:composition}), Eqs.~(\ref{eq:free_A}) and (\ref{eq:free_B})  can be developed;
%
\begin{equation}
F{\simeq}F_A+\frac
{{x_A}^2+\left({\alpha}{\cdot}\grave{E}_{\rm{comp+}}^{\rm{Proj}}-1\right)x_A-{\alpha}{\cdot}\grave{E}_{\rm{comp+}}^{\rm{Proj}}}
{\alpha\cdot{\langle}q_{\rm{comp}}\rangle^{(+G)}_{\rm{comp}}}.
\label{eq:develop_FA}
\end{equation}
\begin{equation}
F{\simeq}F_B-\frac
{{x_B}^2+\left({\alpha}{\cdot}\grave{E}_{\rm{comp-}}^{\rm{Proj}}-1\right)x_B-{\alpha}{\cdot}\grave{E}_{\rm{comp-}}^{\rm{Proj}}}{\alpha\cdot{\langle}q_{\rm{comp}}\rangle^{(-G)}_{\rm{comp}}}.
\label{eq:develop_FB}
\end{equation}
\begin{equation}
\alpha=\frac{{\sqrt{2\pi}{\langle}q_{\rm{comp}}\rangle_{2G}}}{{k_{\rm{B}}T}}.
\label{eq:develop_alpha}
\end{equation}
%
  Note that ${\langle}q_{\rm{comp}}\rangle^{(-G)}_{\rm{comp}}\leq{0}$ because we define ${\langle}I\rangle_1 = 0$ after Eq.~\ref{eq:composition}. These development successfully shows that Helmholtz free energy $F$ can be described only by GPS (projection energy and ${\langle}q_{\rm{comp}}\rangle^{({\pm}G)}_{\rm{comp}}$)  even though $F$ has entropy term depending on all possible microscopic states.  Here, we have two representation based on A-rich and B-rich system. Difference of these value comes from approximation of density of states (explained above). However, these two types of representation are both efficient even though their differences. It is because $F$ has introduced for value comparison of stability, and Eqs.~(\ref{eq:develop_FA}) and (\ref{eq:develop_FB}) respectively tells stability with comparison to A and B unary system. In addition to this, this two representation indicate phase separation between disorder states like fig.~\ref{fig.image_F}.The connection of free energy, composition and temperature shows in appendix C with image graph.
 \\
 \begin{figure}[t]
 \centering
\includegraphics[width=8cm]{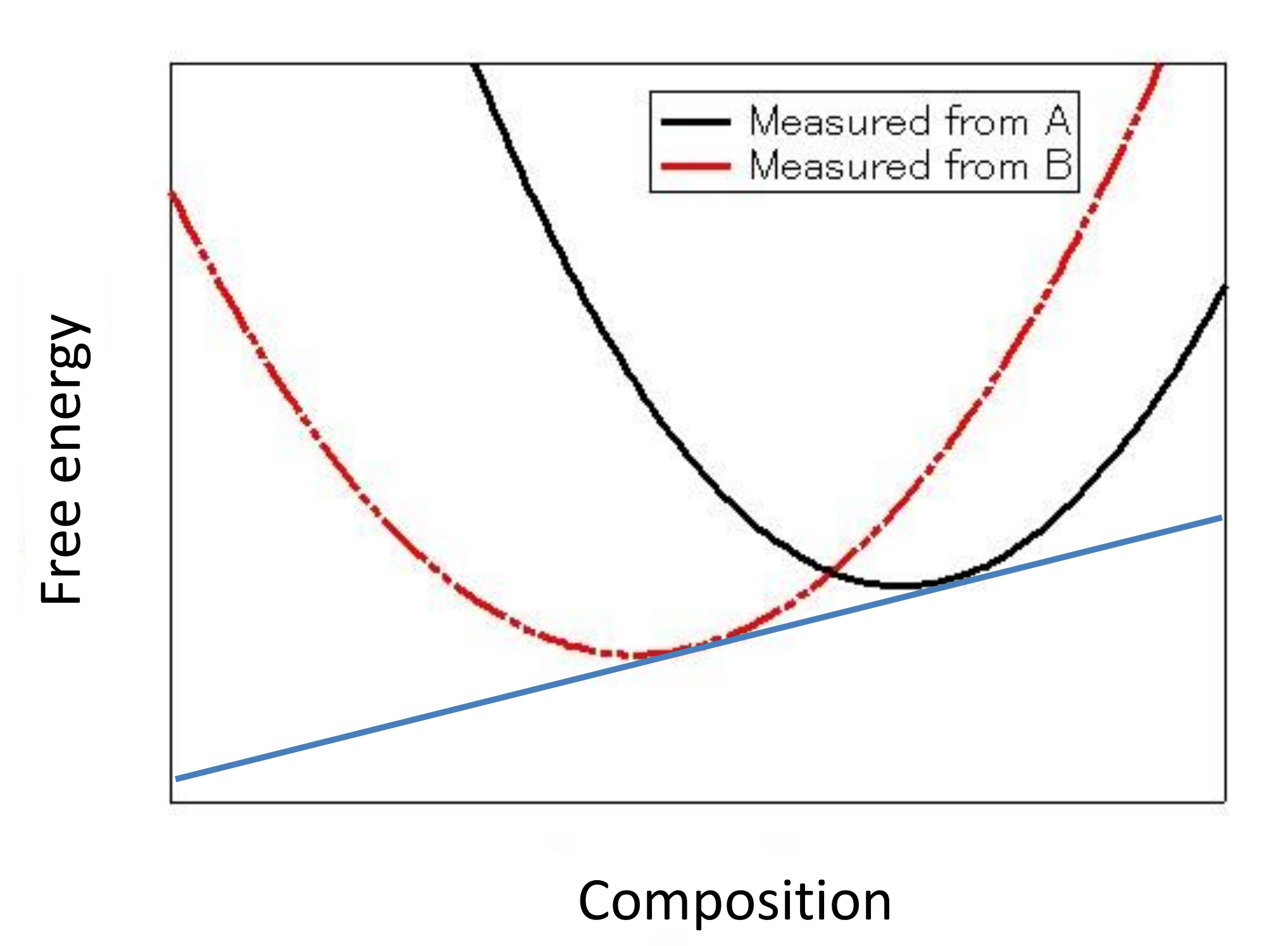}
\caption{Image graph of free energy and composition. Blue line shows that two representation of free energy reveal the phase separation between disordered states.}
\label{fig.image_F}
\end{figure}
 \ In multi-component system, we can use this approach; a single state can describe canonical average of $q_r$. When we consider $N$-component system, we separate basis functions $\{q_1,q_2,...q_s\}$ into two part; part of composition $\{v_1,v_2,...,v_{N-1}\}$ and part of position $\{w_1,w_2,...,w_{s-N+1}\}$. With a change of basis,  we set each function of composition describing composition of some element. With this setting our present approach gives a single microscopic state describing composition for one element. However for getting free energy, we need composition of all elements, thus we need ($N-1$) microscopic states. They are our future works.
 \section{\label{sec:level4}conclusion}
In present study, we confirm that composition unfixed practical system is more similar to random matrix than composition-fixed system in terms of vanishment of interdependence of $q_r$. This result suggest to us that our canonical approach should be extended to semi-grand canonical one. Then, we successfully discover a single special microscopic state (GPS) in binary system only from the information of spatial constraint. This GPS gives us composition in the system. Furthermore, GPS gives us Helmholtz free energy measured from unary system, which has not been described by a single state.

\section*{Acknowledgement}
This work was supported by a Grant-in-Aid for Scientific Research (16K06704), and a Grant-in-Aid for Scientific Research on Innovative Areas “Materials Science on Synchronized LPSO Structure” (26109710) from the MEXT of Japan, Research Grant from Hitachi Metals·Materials Science Foundation, and Advanced Low Carbon Technology Research and
 Development Program of the Japan Science and Technology Agency (JST)."

%

\setcounter{equation}{0}
\setcounter{figure}{0}
\setcounter{table}{0}
\setcounter{section}{0}
\appendix
\renewcommand{\theequation}{A\arabic{equation}}
\renewcommand{\thefigure}{A\arabic{figure}}

\section {Effect of composition}
\subsection{ Composition-fixed/unfixed method}
 In this section, we explain two method to set practical system in random matrix research; composition-fixed and composition-unfixed method. In the fixed method, we set the atom under fixed composition through MC simulation. 
\\
\ In the unfixed method, we consider local system, $l$, in ideally large composition-fixed system, $f$. In this situation, this local system should have composition-fluctuation; probability $P_x$ that local system has A-composition $x$ in binary system  is theoretically given by binomial distribution:
\begin{equation}
P_x=_{N_l}\mathrm{C}_{xN_l}{x_f}^{xN_l}(1-x_f)^{N-xN_l}.
\end{equation}
 Here, $N_l$ is number of local system, $x_f$ is composition of large system. In $m$ times sampling, $m{\cdot}P_x$ is the times sampled at composition $x$. Therefore, for getting matrix from local system, we set the atom $m{\cdot}P_x$ times under composition $x$, and do MC simulation. In this setting, we can consider local composition fluctuation of the large system. \\
\ In present study, we set equiatomic large box in unfixed method and consider 1152-atom local system (i.e. $x_f=0.5$ and $N_l=1152$)  for comparison to equiatomic fixed system. Figure~\ref{fig.com_sampling} shows connection of composition and sampling number in unfixed method. 
 
 \begin{figure}[h]
\includegraphics[width=8cm]{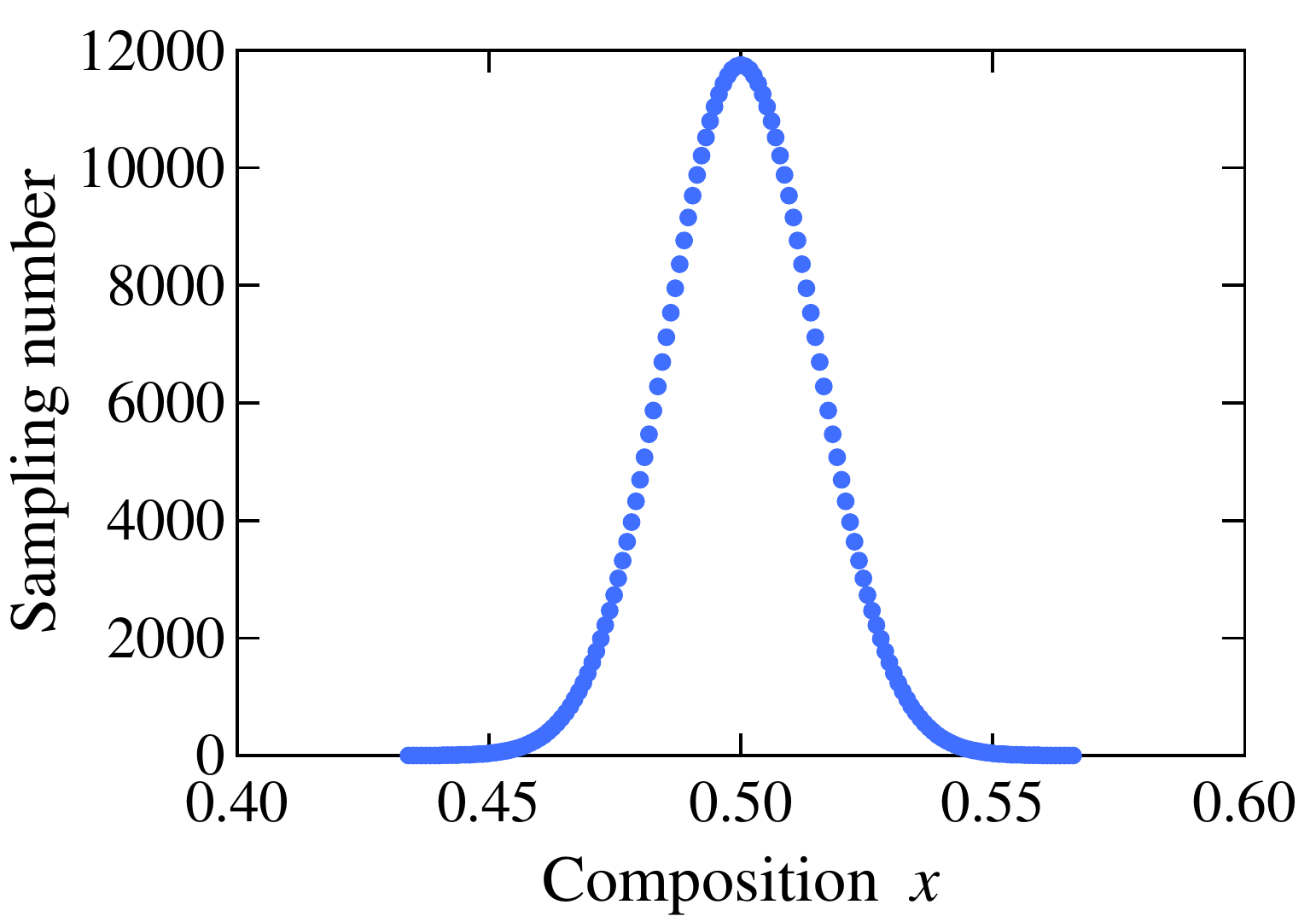}
\caption{Sampling number for composition in composition-unfixed practical system. No plot point along composition does not be sampled in this trial because of the number of sampling.}
\label{fig.com_sampling}
\end{figure}
\subsection{Analytical representation of GPS}
\ In this section, we show that GPS can be represented without any simulations. First, average of $q_r$ is the function of composition,$\overline{q_r}=(2x_A-1)^{n_p}$\cite{hensa}\cite{SQS}, where $n_p$ is the number of lattice points per one cluster. Probability of composition, $P_x$ is also given by binomial distribution leading to Gaussian distribution;
\begin{eqnarray}
P_x&=&_{N}\mathrm{C} _{xN}\left(\frac{1}{2}\right)^N\\
&\simeq&\sqrt\frac{2N}{{\pi}}\exp\left\{-2N\left(x-\frac{1}{2}\right)^2\right\}
\end{eqnarray}
  Hence, when we decide what kind of and how many clusters we think, GPS can be represented analytically.\\
\renewcommand{\theequation}{B\arabic{equation}}
  \section{Another projection state}
  In reference \cite{EMRS2},we have successfully introduce concept of projection state and energy;
     \begin{equation}
Q_r(T)\simeq{\langle}q_r\rangle_{1}-\sqrt{\frac{\pi}{2}}{\langle}q_r\rangle_{2}\cdot\frac{E_r^{\rm{proj}}-{\langle}E{\rangle}_1}{k_{\rm{B}}T}.
\label{eq:Projection_GCE7}
\end{equation}
 \begin{equation}
E_{r}^{\rm{proj}}=\sum_{t=1}^{s} {\langle}E{\mid}q_t\rangle{\langle}q_t\rangle_r^{(+)},
\label{eq:Projectionenergy_CE73}
\end{equation}
  \\
\ In this approach, we have considered partial average of $q_r\ge{\langle}q_r{\rangle}_1$. However, when we consider that the approach relies on the approximation that density of state on configuration space for non-interacting system can be regarded as multidimensional Gaussian distribution, it is more natural that we consider another partial average; average under $q_r\leq{\langle}q_r\rangle_1$.
\\
\ Thus, we express $E_r^{\rm{proj}}$ under $q_r\geq{\langle}q_r\rangle_1$ as  $E_{r+}^{\rm{proj}}$ below and introduce new projection energy  $E_{r-}^{\rm{proj}}$, partial average under $q_r\leq{\langle}q_r\rangle_1$;
 \begin{eqnarray}
E_{r-}^{\rm{proj}}-{\langle}E\rangle_1 &=& 2\int_\infty^\infty{\int}^{{\langle}q_r\rangle_1}_{q_r^{min}}E{\cdot}g(E,q_r)dq_rdE\nonumber\\
&=&\sqrt\frac{2}{\pi}\Gamma_{12}{{\langle}q_r\rangle_2}^{-1}\left\{\exp\left[-\left(\frac{q_r^{min}}{\sqrt{2}{\langle}q_r\rangle_2}\right)\right]^2-1\right\}.\nonumber\\
\label{eq:projection2}
\end{eqnarray}
With taking the limit of N, we can develop the equation same as $E_{r+}^{\rm{proj}}$;
   \begin{equation}
Q_r(T)\simeq{\langle}q_r\rangle_{1}+\sqrt{\frac{\pi}{2}}{\langle}q_r\rangle_{2}\cdot\frac{E_{r-}^{\rm{proj}}-{\langle}E{\rangle}_1}{k_{\rm{B}}T}.
\label{eq:Projection_GCE2}
\end{equation}
 It is described by a single state, \{${\langle}q_1\rangle_r^{(-)},{\langle}q_2\rangle_r^{(-)},...{\langle}q_s\rangle_r^{(-)}$\}, where $\langle\cdot\cdot\cdot\rangle_r^{(-)}$ is a partial average under $q_r\le{\langle}q_r{\rangle}_1$. Consequently, we get two representation of $Q_r$;
  \begin{equation}
Q_r(T)\simeq{\langle}q_r\rangle_1\mp\sqrt{\frac{\pi}{2}}{\langle}q_r\rangle_2\cdot\frac{E_{r\pm}^{\rm{proj}}-{\langle}E\rangle_1}{k_{\rm{B}}T}.
\label{eq:Projection_CE5}
\end{equation}
 Two values of $Q_r$ can be different due to the approximation. When the density of states completely match Gaussian distribution, these values are same. 
 \section{ The landscape of free energy, composition and temperature}
 \renewcommand{\thefigure}{C\arabic{figure}}
 \setcounter{figure}{0}
 Figure.~\ref{fig.image_F_with_T} shows connection of free energy, composition and temperature with Eq.~(\ref{eq:develop_FA}) under constant projection energy. From Eq.~(\ref{eq:develop_FA}) and Eq.~(\ref{eq:develop_FB}) we realize that the system take minimum free energy at equiatomic composition with taking limit of thermodynamics; $T\to\infty$. It matches conventional thermodynamics. 
   \begin{figure}[h]
\centering
\includegraphics[width=8cm]{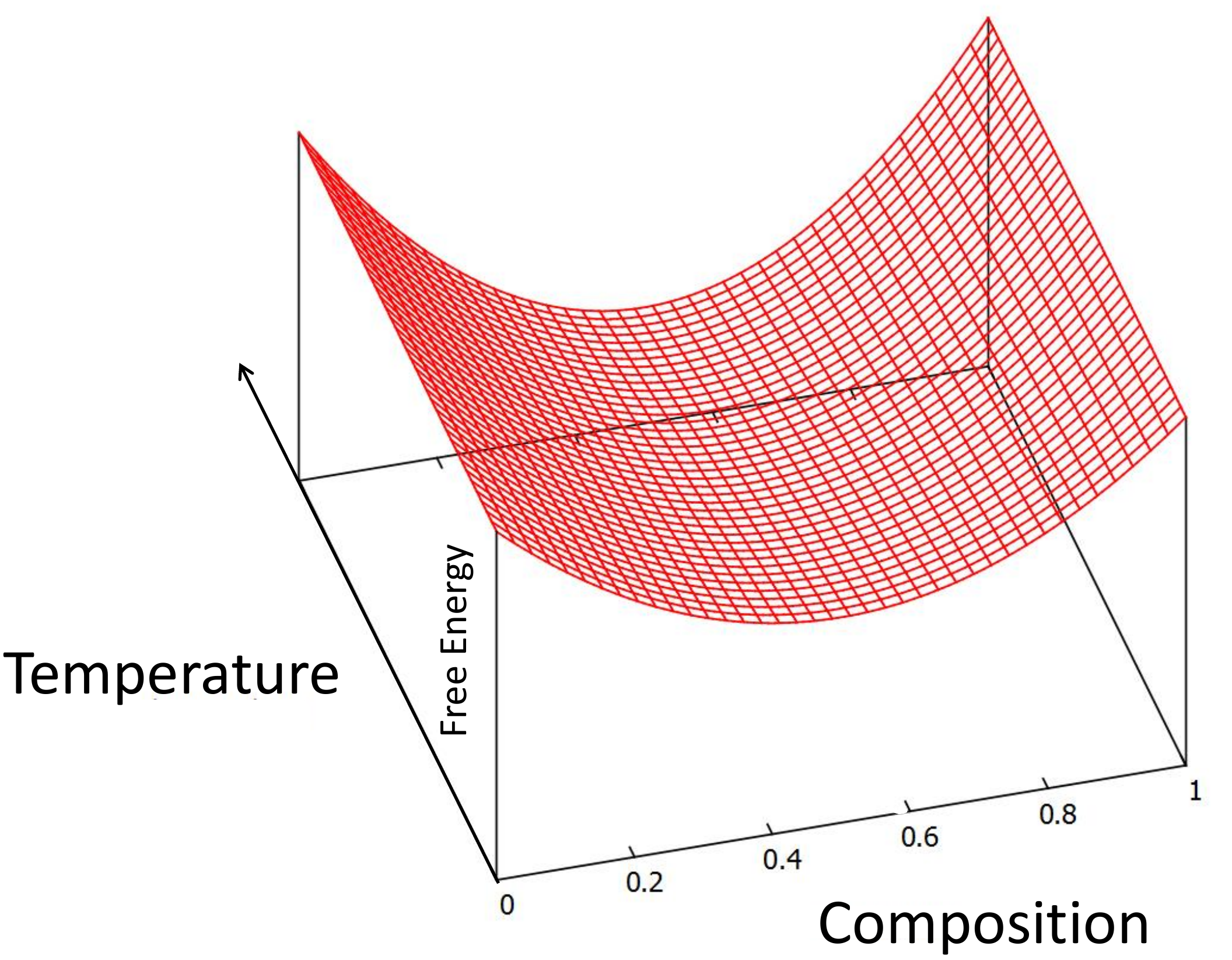}
\caption{Image graph of free energy and composition and temperature under constant projection energy. Temperature and free energy is normalized.}
\label{fig.image_F_with_T}
\end{figure}


\begin{thebibliography}{999}
\bibitem{EntropicSampling} J. Lee, Phys. Rev. Lett. {\bf71}, 211 (1993).
  \bibitem{MonteCarlo1} Metropolis N, Rosenbluth A W, Rosenbluth M N, Teller A H and Teller E, J. Chem. Phys. {\bf21} 1087 (1953).
  \bibitem{MonteCarlo2} A. M. Ferrenberg and R. H. Swendsen, Phys. Rev. Lett. {\bf63}, 1195 (1989).
  \bibitem{CVM1} Kikuchi R, Phys. Rev. {\bf81} 988 (1951).
  \bibitem{CVM2} Kikuchi R, J. Chem. Phys. {\bf60} 1071 (1974).
  \bibitem{Frankel} B. J. Alder, S. P. Frankel and V. A. Lewison, J. Chem. Phys. {\bf23}, 417 (1955)
 \bibitem{EMRS1} K. Yuge, J. Phys. Soc. Jpn. {\bf84}, 084801 (2015).
 \bibitem{EMRS2} K. Yuge, J. Phys. Soc. Jpn. {\bf85}, 024802 (2016).
 \bibitem{EMRS3}K. Takeuchi, R. Tanaka and K. Yuge, J. Phys.:Condens. Matter {\bf27}, 385201 (2015).
 \bibitem{CEM1} J. W. Connolly and A. R. Williams, Phys. Rev. {\bf27}, 5169 (1983).
 \bibitem{CEM2} J. M. Sanchez, F. Ducastelle and D.Gratias, Pysica {\bf128A}, 334 (1984).
 \bibitem{CEM3} T. Mouri and Y. Chen, J.Japan Inst. Metals, {\bf68}, 996 (2004).
\bibitem{EMRS4} K. Yuge,  T. Kishimoto and K. Takeuchi, Trans. Mat. Res. Soc. Jpn. {\bf{41}} 213 (2016).
\bibitem{Marchenko} V.A. Marchenko and L.A. Pastur, Math. USSR Sb. {\bf1} 457 (1967).
\bibitem{F_vib}D. de Fountain, in Solid State Physics, ed. H. Ehrenreich and D.Turnbull (Academic Press, Cambridge, MA, 1994) Vol. 47, p. 33.
\bibitem{hensa} S.H. Wei, L. G. Ferreira, J. E. Bernard, and A. Zunger, Phys. Rev. B {\bf{42}}, 9622 (1990).
\bibitem{SQS}A. Zunger, S.-H. Wei, L. G. Ferreira, and J. E. Bernard, Phys. Rev.Lett. {\bf65}, 353 (1990).
\end{thebibliography}
\end{document}